%Paper: hep-th/9503105
%From: LABASTIDA@GAES.USC.ES
%Date: Thu, 16 Mar 1995 12:59:33 WST
%Date (revised): Tue, 21 Mar 1995 09:07:44 WST

%\draft
\input phyzzx.tex

\tolerance=500000 \overfullrule=0pt
\def\np{Nucl. Phys.}
\def\pl{Phys. Lett.} \def\pre{Phys. Rep.} 
  \def\cmp{Comm. Math. Phys.}
      \def\topo{Topology}   \def\jmp{J. Math. Phys.} \def\jgp{J.
Geom. Phys.}

\def\too{\longrightarrow}

\tolerance=500000 \overfullrule=0pt

\pubnum={US-FT-3/95\cr
hepth@xxx/9503105}
%\pubnum={US-FT-3/95}
\date={March, 1995}
\pubtype={}
\titlepage \title{A TOPOLOGICAL LAGRANGIAN FOR MONOPOLES ON FOUR-MANIFOLDS}
\author{J.M.F. Labastida\foot{E-mail: LABASTIDA@GAES.USC.ES} and  M.
Mari\~no} \address{Departamento de F\'\i sica de Part\'\i culas\break
Universidade de Santiago\break E-15706 Santiago de Compostela, Spain}
\abstract{We present a topological quantum field theory which corresponds to
the moduli problem associated to Witten's monopole equations for
four-manifolds. The construction of the theory is carried out in purely
geometrical terms using the Mathai-Quillen formalism, and the corresponding
observables are described. These provide a rich set of new topological
quantites.}

\endpage
\pagenumber=1
\sequentialequations

In a recent work Witten
\REF\mfm{E. Witten, ``Monopoles and four-manifolds",
preprint IASSNS-HEP-94-96, hep-th/9411102, 1994} [\mfm] has shown that
Donaldson theory  \REF\don{S. Donaldson\journal\topo&29(90)257} \REF\tqft{E.
Witten\journal\cmp&117(88)353} [\don,\tqft] with gauge group $SU(2)$ is
equivalent to a new moduli problem which involves an abelian Yang-Mills
connection and a spinor coupled in a pair of ``monopole equations".  This
result is a consecuence of previous work  on $N=2$ and $N=4$ Yang-Mills
theory \REF\ws{N. Seiberg and E. Witten
\journal\np&B426(94)19 \journal\np&B431(94)484 } \REF\wv{C. Vafa and E.
Witten \journal\np&B431(94)3 } \REF\wjmp{E. Witten\journal\jmp&35(94)5101}
[\wjmp,\ws,\wv]. The equivalence discovered by Witten is very powerfull and
allows to write explicit expressions for Donaldson polynomials. An inmediate
task which arises from his work is the search for a topological quantum field
theory  related to the new moduli problem presented in [\mfm]. The
observables of such topological quantum field theory could provide new
topological invariants which could contain important topological information.

The aim of this paper is to construct the topological quantum field theory
corresponding to  the new moduli problem proposed in [\mfm].  This will be
done using  \REF\mq{V. Mathai and D. Quillen\journal\topo&25(86) 85} the
Mathai-Quillen formalism [\mq]. The resulting theory turns out to be an
abelian Donaldson-Witten theory, which as it is widely known can be obtained
from the twisting of $N=2$ Yang-Mills theory, coupled to a twisted version of
the $N=2$ hypermultiplet  \REF\fay{P. Fayet\journal\np&B113(76)135}
\REF\son{M.F. Sohnius\journal\np&B138(78)109} \REF\hst{P.S. Howe, K.S. Stelle
and P.K. Townsend\journal\np&B214(83)519} [\fay,\son,\hst]. The resulting
type of topological model has been studied previously in \REF\alab{M. Alvarez
and J.M.F. Labastida\journal\pl&B315(93)251}  \REF\alabas{M. Alvarez and
J.M.F. Labastida\journal\np&B437(95)356} [\alab, \alabas]. Related
topological quantum field theories have been analyzed in \REF\ans{D. Anselmi
and P. Fre\journal\np&B392(93)401\journal\np&B404(93)288
\journal\np&B416(94)255} [\ans], and their connection to the moduli problem
presented in [\mfm] has been recently  considered in \REF\ansdos{D. Anselmi
and P. Fre, ``Gauged hyperinstantons and monopole equations", HUTP-94/A041,
hep-th/9411205, 1994} [\ansdos].

The Mathai-Quillen formalism allows one to construct the action of a
topological quantum field theory starting from moduli problems formulated in
purely geometrical  terms. Moduli problems are often stated in the following
form: given a moduli space ${\cal M}$ and a vector bundle over ${\cal M}$,
${\cal V}$, one defines the basic equations of the problem as sections of
this vector bundle. Typically one is interested in computing the Euler
characteristic of this bundle, or, equivalently, its Thom class. In the case
at hand, because of the gauge symmetry of the theory, one also has the action
of a group ${\cal G}$ on both, the manifold ${\cal M}$ and the vector
bundle.  Rather than compute the Euler characteristic of the bundle itself
one wants to get rid of the gauge degrees of freedom and compute the Euler
characteristic of the quotient bundle obtained  ``dividing by ${\cal G}$":
${\cal V} /{\cal G} \too {\cal M}/{\cal G}$. In the same way, the section
$s:{\cal M} \too {\cal V}$ is taken to be gauge-equivariant and  hence one
can  define the associated section ${\hat s}:{\cal M}/{\cal G} \too {\cal V}
/{\cal G}$.

As in the Donaldson-Witten case, for monopoles on four-manifolds the vector
bundle is trivial and can be written as ${\cal V} = {\cal M} \times {\cal
F}$, where ${\cal F}$ is the fibre on which a ${\cal G}$-invariant metric is
defined. If one considers the moduli space ${\cal M}$ as a principal bundle
with group ${\cal G}$ the quotient bundle is  the associated vector bundle
${\cal E}={\cal M} {\times _{\cal G}} {\cal F}$. This is the situation
analyzed in  \REF\aj{M.F. Atiyah and L. Jeffrey\journal\jgp&7(90)119} [\aj],
where the results are particularized for Donaldson-Witten theory. A more
general situation, involving non-trivial vector bundles, is considered in
\REF\cmrym{S. Cordes, G. Moore and S. Rangoolam, ``Large N 2D Yang-Mills
Theory and Topological String Theory", hep-th/9402107, 1994} \REF\cmrlect{S.
Cordes, G. Moore and S. Rangoolam, Proceedings of the 1994 Les Houches
Summer School} [\cmrym, \cmrlect].

To define the Mathai-Quillen form of the associated bundle ${\cal E}$ one
needs a connection on it. If the space ${\cal M}$ has a ${\cal G}$-invariant
metric defined on it there is a natural way to construct it as follows [\aj]:
consider on the principal bundle ${\cal M} \too {\cal M}/{\cal G}$ the
connection defined by declaring the horizontal subspaces to be the orthogonal
ones to the vertical subspaces. The latest are just the gauge orbits given by
the action of the group ${\cal G}$. This connection on the principal bundle
${\cal M}$ induces a connection on the associated bundle ${\cal E}$ in the
standard way, and this is just the connection that one needs in the
construction of the topological lagrangian. Notice, however, that in more
general situations (mainly when the vector bundle of the moduli problem is
not a trivial one, as it happens in topological string theory) one must add
another connection to the previous one [\cmrym, \cmrlect].

As in the Donaldson-Witten theory, we will use the Cartan model for the
equivariant cohomology which gives the BRST symmetry of the theory. Hence we
will deal with the Cartan model of the Mathai-Quillen form. This is an
equivariant differential form of the fibre ${\cal F}$ which can be written
as:
$$
 U={\rm e}^{-|x|^2} \int D \chi {\rm exp}\Big( {1 \over 4} \langle
\chi , \Omega \chi \rangle  + i\langle dx , \chi \rangle \Big).
\eqn\mathai
$$
In this expression, $x$ denotes a (conmuting) vector coordinate for the
fibre ${\cal F}$,  $\chi$ a Grassman coordinate and the bracket a  ${\cal
G}$-invariant metric on ${\cal F}$. $\Omega$ is the universal curvature
which acts on the fibre according to the action of the group ${\cal G}$.
Now, in order to obtain a differential form on the base space ${\cal
M}/{\cal G}$ we must use the Chern-Weil homomorphism which has the effect of
substituting $\Omega$ by the actual curvature on ${\cal M}$ and thus gives a
basic differential form on ${\cal M} \times {\cal F}$. However, in the
Cartan model, due to the relation between the Cartan model and the Weil
model for equivariant cohomology, one needs to make an horizontal projection
in order to obtain a closed form on ${\cal E}$. In other words, the
differential form on ${\cal M} \times {\cal F}$ must be evaluated on the
horizontal subspace of ${\cal M}$. Once we do that, we have a form on ${\cal
E}$ which descends to a form on ${\cal M}/{\cal G}$ by simply taking the
pullback by the section ${\hat s}$. This has the effect of substituting the
coordinate $x$ by the section ${\hat s}$.

Let us describe in detail how to construct the connection on ${\cal M}$ and
how to enforce the horizontal projection. The gauge orbits are given by the
vertical tangent space on the principal bundle with group ${\cal G}$, which
is given by a map from the Lie algebra of the group ${\cal G}$, which we
denote by  ${\rm Lie}({\cal G})$, to the tangent space to ${\cal M}$,
$$
C:{\rm Lie}({\cal G})  \too T{\cal M}.
\eqn\ope
$$
We will assume that
both ${\rm Lie}({\cal G})$ and ${\cal M}$ are provided with metrics (in the
case of ${\rm Lie}({\cal G})$ this is simply an appropriate generalization of
the Cartan-Killing form) so we can consider the adjoint operator
$C^{\dagger}$ and the operator $R=C^{\dagger}C$. The connection one-form is
given by [\aj],
$$
\Theta = R^{-1} C^{\dagger}.
\eqn\lima
$$
As the
Cartan representative acts on horizontal vectors, we can write the curvature
as
$$
\Omega = d \Theta =R^{-1}dC^{\dagger}.
\eqn\curv
$$
Now, to enforce
the horizontal projection we should have to integrate over the vertical
degrees of freedom which amounts to an integration over the Lie group.
Alternatively, we can introduce a ``projection form" [\cmrlect] which,
besides of projecting on the horizontal direction, automatically involves the
Weil homomorphism which substitutes the universal curvature by the actual
curvature on the bundle \curv. The projection form also allows to write the
correlation functions on the quotient moduli space ${\cal M}/{\cal G}$ as
integrals over ${\cal M}$. Taking into account all these facts, and after
some suitable manipulations, we obtain the folowing expression for the Thom
class of the bundle ${\cal E}$:
$$
\int  D\eta D\chi D\phi D \lambda \,
{\rm exp}\Big( -|s|^2 + {1 \over 4} \langle \chi , \phi \chi \rangle_{g} +
i\langle ds , \chi \rangle +i{\langle dC^{\dagger}, \lambda \rangle }_{g}-
i{\langle \phi,R \lambda \rangle }_{g} +i\langle C^{\dagger} \theta , \eta
\rangle_{g} \Big).
\eqn\mqaj
$$
Here, $\phi$, $\lambda$ are conmuting Lie
algebra variables and $\eta$ is a Grassmann one. The variables $(P, \theta)$
(the first one is conmuting and present in $s$, the second one is Grassmann)
are the usual superspace coordinates for the integration of differential
forms on ${\cal M}$. The bracket with the subscript $g$ is the
Cartan-Killing form of ${\rm Lie}({\cal G})$. This expression is to be
understood as a differential form on ${\cal M}$ which when integrated out
with the measure $DP D\theta$ gives the Euler characteristic of ${\cal E}$.

Let us consider now the moduli problem of monopoles on four-manifolds
proposed in [\mfm]. Let $X$ be a spin four-manifold, endowed with a
Riemannian metric $g_{ij}$. Denote by $S^{+}$ and $S^{-}$ the positive and
negative chirality spin bundles on $X$, respectively. Consider in addition a
complex line bundle $L$ with an associated $U(1)$ connection. Let ${\cal A}$
denote the moduli space of these abelian connections, and $\Gamma (X, S^{+}
\otimes L)$ the sections of the product bundle $S^{+} \otimes L$, {\it
i.e.}, positive chirality spinors taking values in $L$. The moduli space of
our problem is thus ${\cal M}={\cal A} \times \Gamma (X, S^{+} \otimes L)$.
The vector bundle over ${\cal M}$ is a trivial one with fibre  ${\cal
F}=\Omega^{2,+} (X) \oplus \Gamma (X, S^{-} \otimes L)$, where the first
factor denotes the self-dual differential forms of degree $2$ on $X$. As in
the Donaldson-Witten case, the group ${\cal G}$ is the group of gauge
transformations of the principal $U(1)$-bundle associated to the connection
$A$, whose action on the moduli space is given locally by:
$$
\eqalign{
&g^{*}(A_i)=A_{i}+ig^{-1}\partial_{i}g,\cr &g^{*}(M_{\alpha})=g
M_{\alpha},\cr }
\eqn\gauge
$$
where $M \in \Gamma (X, S^{+} \otimes L)$
and $g$ takes values in $U(1)$ . The group of gauge transformations also
acts on the fibre ${\cal F}$, but we must use $g^{-1}$, as  the construction
of an associated vector bundle imposes. Also notice  that there is no action
on the factor $\Omega^{2,+} (X)$, for the group is abelian. The Lie algebra
of the group ${\cal G}$ is just ${\rm Lie}({\cal G})=\Omega^{0}(X)$, as the
Lie algebra of $U(1)$ is  ${\bf R}$. Now we need metrics on both the moduli
space and the vector bundle. The tangent space to the moduli space at the
point $(A,M)$ is just $T_{(A,M)}{\cal M}=T_{A}{\cal A} \oplus T_{M}\Gamma
(X, S^{+} \otimes L)=\Omega^{1}(X) \oplus \Gamma (X, S^{+} \otimes L)$, for
$\Gamma (X, S^{+} \otimes L)$ is a vector space. The metric on ${\cal M}$ is
given by:
$$
\langle (\psi, \mu), (\theta, \nu) \rangle=\int_{X} \psi
\wedge *\theta +{1 \over 2} \int_{X} e ({\bar \mu}^{\alpha} \nu_{\alpha}+ \mu
_{\alpha} {\bar \nu}^{\alpha}),
\eqn\pina
$$
where $e=\sqrt g$. For
spinors we use the  following notation. If $\mu_{\alpha}=(a,b)$,
$\bar\mu^{\alpha}$ is chosen as $\bar\mu^{\alpha}=(a^*,b^*)$. Notice that
throughout this work the signature of the metric $g_{ij}$ is Euclidean.
Spinor indices al lowered and rised using the invariant tensor
$C_{\alpha\beta}$ as in    \REF\roc{S.J. Gates, M.T. Grisaru, M. Ro\v cek
and W. Siegel, ``Superspace", Benjamin, 1983} [\roc] ($C_{\alpha\beta}$ are
the entries of  the Pauli matrix $\sigma^2$, for example,
$\bar\mu_{\alpha}=\bar\mu^{\beta}C_{\beta\alpha}=(ib^*,-ia^*)$). An
expression analogous to \pina\ gives the inner product on the vector bundle
${\cal V}$. Notice that we are considering both $S^{+} \otimes L$ and
$S^{-} \otimes L$ as {\it real} vector spaces of dimension four.

Let us now  introduce the section associated to  the ``monopole equations"
in [\mfm]. Recall that the product of two spinors  can be decomposed in
terms of a $0$-form and a self-dual $2$-form. This allows one to write:
$$
s(A,M)=\Big({1 \over \sqrt 2} (F^{+}_{\alpha \beta}+{i\over 2} {\overline
M}_{(\alpha}  M_{\beta )}), D_{\alpha {\dot \alpha} }M^{\alpha}\Big).
\eqn\rosa
$$
In this expression $D_{\alpha {\dot \alpha}}$ is the
Dirac operator. In our notation, $D_{\alpha {\dot \alpha}} M_\beta =
\sigma^j_{\alpha {\dot \alpha}} (\partial_j+iA_j)M_\beta$ where the matrices
$\sigma^j$ are $ \sigma^j = ({\bf 1},i\sigma^1,i\sigma^2,i\sigma^3)$, being
$\sigma^1$, $\sigma^2$ and $\sigma^3$ the Pauli matrices. In \rosa\
$F^{+}_{\alpha \beta}$ is the self-dual part of the gauge field-strength:
$$
F^{+}_{\alpha \beta}= (p^+(F))_{\alpha\beta}=
C^{\dot\alpha\dot\beta}(\sigma^i)_{\alpha\dot\alpha}(\sigma^j)_{\beta\dot\beta}
{1\over 2}(F_{ij}-{1\over 2e}\epsilon_{ij}{}^{kl}F_{kl}),
\eqn\proy
$$
being
$C^{\dot\alpha\dot\beta}$ the matrix $-\sigma^2$, and $\epsilon_{ijkl}$ the
totally antisymmetric tensor density. In \proy\ $p^+(Z)$ denotes  the
projection of a two-form $Z$ into its self-dual part. The factor
$1/{\sqrt{2}}$ in \rosa\ has been introduced for convenience as will be
explained below.

Once  the underlying geometry of the model has been presented we will
construct the topological lagrangian from the general expression \mqaj. First
of all we must describe explicitly the tangent vertical space by means of the
operator $C$. This is simply obtained from \gauge, and it reads:
$$
C(\epsilon)=(-d\epsilon ,i\epsilon M) \in \Omega^{1}(X) \oplus \Gamma (X,
S^{+} \otimes L),\,\,\,\,\ \epsilon \in \Omega^{0}(X).
\eqn\flor
$$
To obtain the adjoint of this operator we must use the Cartan-Killing form on
${\rm Lie}({\cal G})=\Omega^{0}(X)$, which is just the usual product of
differential forms on $X$. One finds:
$$
C^{\dagger}(\psi,\mu)=-d^{*}\psi +
{i \over 2}({\bar \mu}^{\alpha}M_{\alpha}-{\overline
M}^{\alpha}\mu_{\alpha}),
\eqn\marga
$$
and hence the operator $R$ is
given by
$$
R=d^{*}d+ {\overline M}^{\alpha}M_{\alpha}.
\eqn\tren
$$
Another operator we need to write the lagrangian is $ds:T_{(A,M)}{\cal M}
\too {\cal F}$. To obtain it we must linearize the monopole equations. The
result is:
$$
ds(\psi,\mu)= \Big( {1 \over \sqrt 2}\big((p^+(d\psi))_{\alpha
\beta}+{i\over 2}({\overline M}_{(\alpha} \mu _{\beta )}+{\bar \mu}_{(\alpha}
M_{\beta )})\big), D_{\alpha {\dot \alpha} } \mu^{\alpha}+i\psi _{\alpha
{\dot \alpha}}M^{\alpha}  \Big),
\eqn\disco
$$
where $p^+$ is the
projection defined in \proy. The maps $C$ and $ds$ are important because
they give the instanton deformation complex:
$$
0 \too \Omega^{0}(X)
\buildrel C \over \too \Omega^{1}(X) \oplus \Gamma (X, S^{+} \otimes L)
\buildrel ds \over \too
 \Omega^{2,+} (X) \oplus \Gamma (X, S^{-} \otimes L) \too 0.
\eqn\file
$$
The index of this complex is precisely minus the dimension of the tangent
space to the zero locus of the section ${\hat s}$, which is the moduli space
of solutions to the monopole equations modulo gauge transformations. These
two operators are also the operators appearing in the fermion kinetic terms
of the lagrangian, as one can see in \mqaj , and the referred index computes
the difference of zero modes of the fermion fields. To obtain the index of
this complex, notice that we can drop out the terms of order zero of the
involved operators, for their leading symbol is not changed. In this way we
obtain an equivalent complex which factorizes into the complex for the Dirac
operator (the leading term for $ds$) and the complex
$$
0  \too
\Omega^{0}(X) \buildrel d \over \too \Omega^{1}(X)  {\buildrel p^{+}d \over
\too} \Omega^{2,+} (X) \too 0
\eqn\filedos
$$
where $p^{+}$ is the projection into the sel-dual part defined in \proy.
The index we are looking for is simply the index of the second complex minus
the index of the Dirac operator multiplied by two. As $X$ is
four-dimensional, we obtain for the last one $c_{1}(L)^2-p_{1}(X)/12$.
For the second complex, the index is
$b_{0}-b_{1}+b_{2}^{+}$, which can be written as $(\chi+\sigma)/2$, where
$\chi$ is the Euler characteristic of $X$ and $\sigma=b_{2}^{+}-b_{2}^{-}$
its signature. Now, using the Hirzebruch signature formula, we have
$\sigma=p_{1}(X)/3$. Taking all this into account, we obtain [\mfm]:
$$
{\rm index}\,\, T={2\chi + 3 \sigma \over 4}- c^{2}_{1}(L).
\eqn\leon
$$

In order to write the topological quantum field  theory associated to the
moduli problem we must indicate the field content and the topological
symmetry. These are determined by the geometrical structure we have been
developing. For the moduli space we have conmuting fields $P=(A,M) \in {\cal
M}= {\cal A} \times \Gamma (X, S^{+} \otimes L)$, with ghost number $0$ and
their superpartners, representing a basis of differential forms on ${\cal
M}$, $\theta=(\psi, \mu)$, with ghost number $1$. Now, we must introduce
fields for the fibre (corresponding to the $\chi$ variable in \mqaj), which
we denote by $(\chi_{ij}, v_{\dot \alpha}) \in \Omega^{2,+} (X) \oplus \Gamma
(X, S^{-} \otimes L)$, with ghost number $1$. It is also useful in the
construction of the action from gauge fermions to introduce auxiliary
conmuting fields with the same geometrical content, $(H_{ij}, h_{\dot
\alpha})$. The gauge symmetry makes necessary to introduce three fields in
${\rm Lie}({\cal G})$, as we have remarked in writing \mqaj. The field
$\phi$, with ghost number $2$, is a conmuting one. It roughly corresponds
 to the universal curvature and enters in the equivariant cohomology of
${\cal M}$. The fields $\lambda$, $\eta$, with ghost number $-2$ and $-1$,
respectively, come from the projection form, as explained in [\cmrlect]. The
BRST cohomology of the model is:
$$
\eqalign{ &[Q,A_{i}]=\psi _{i},\cr
&\{Q,\psi _{i} \}=\partial_{i}\phi,\cr &[Q, \phi]=0\cr &\{Q,\chi_{ij}
\}=H_{ij}, \cr &[Q,H_{ij}]=0, \cr &[Q, \lambda]=\eta, \cr} \qquad\qquad
\eqalign{&[Q,M_{\alpha}]=\mu _{\alpha}, \cr &\{Q, \mu _{\alpha} \}=-i\phi
M_{\alpha},\cr & \{Q, v_{\dot \alpha} \}=h_{\dot \alpha}, \cr &[Q,h_{\dot
\alpha}]=-i \phi v_{\dot \alpha}, \cr & \{Q, \eta \}=0.\cr}
\eqn\pera
$$
For the fields on the base space the BRST operator is the Cartan
differential for equivariant cohomology $Q=d-{\imath}_{\phi}$, where
${\imath}_{\phi}$ denotes the interior product on differential forms. The
fields $\psi_i$ represent a basis of differential forms and they can be
interpreted formally as $dA_i$ (we can also see them as a basis of tangent
vectors). Notice also that $Q^2 A_i = -{\cal L}_{\phi} A_i$, where ${\cal
L}_{\phi}$ denotes the Lie derivative generated by $\phi$. The same
considerations apply to the fields $M_{\alpha}$, $\mu_{\alpha}$. For the
fibre variables we close the algebra up to a gauge transformation generated
by $-\phi$ (recall that the group acts on the fibre with $g^{-1}$).  We are
now in the position of writing out the action of the theory. Let us
consider first the last five terms in the exponential of the Thom class
\mqaj,
$$
\eqalign{  -&i{\langle \phi, R\lambda \rangle }_{g} =-i\int _{X}
\lambda  \wedge *d^{*}d \phi -i\int _{X}e \phi \lambda{\overline M}^{\alpha}
M_{\alpha},  \cr &i\langle  (\chi , v), ds \rangle = {i \over {\sqrt2}}\int
_{X} \chi \wedge * p^{+}d \psi -{1 \over 2{\sqrt2}}\int _{X}e {\chi}^{\alpha
\beta}({\overline M}_{(\alpha} \mu _{\beta )}+{\bar \mu}_{(\alpha} M_{\beta
)})  \cr  &\,\,\,\,\,\,\,\,\,\,\,\,\,\,\,\,\,\,\,\,\,\,\,
+{i \over 2}\int
_{X} e (\bar v^{\dot \alpha} D_{\alpha {\dot\alpha}} {\mu}^{\alpha}-{\bar
\mu}^{\alpha} D_{\alpha {\dot\alpha}}v^{\dot \alpha}) +{1 \over 2}\int _{X}
e ({\overline M}^{\alpha}{\psi}_{\alpha {\dot \alpha}}v^{\dot \alpha}-{\bar
v}^{\dot \alpha}{\psi}_{\alpha {\dot \alpha}}M^{\alpha}), \cr   &i\langle
C^\dagger (\psi,\mu_{\alpha}) ,  \eta \rangle_g   =-i\int_{X} \eta \wedge *d
\psi -{1\over2} \int _{X} e \eta ({\bar \mu} ^{\alpha} M_{\alpha}-{\overline
M}^{\alpha} \mu_{\alpha}), \cr  &{1 \over 4}{\langle (\chi , v), \phi (\chi
, v) \rangle}_{g} = -{i \over 4}\int _{X} e \phi {\bar v}^{\dot\alpha}
v_{\dot\alpha}, \cr  &i{\langle dC^{\dagger}, \lambda \rangle}_{g}=\int _{X}
e \lambda {\bar \mu}^{\alpha} \mu _{\alpha}.\cr }
\eqn\fresa
$$
The two last terms are obtained as follows. For the term involving ${\langle
(\chi , v), \phi (\chi , v)\rangle}_{g}$ one must take into account the
action of ${\rm Lie}({\cal G})$ on $v_{\dot \alpha}$, which lives in the
fibre: $\phi (v_{\dot \alpha})=-i\phi v_{\dot \alpha}$. On the $2$-forms,
the action of ${\rm Lie}({\cal G})$ is trivial, because the Lie algebra of
$U(1)$ is abelian. To compute $dC^{\dagger}$, which is a $2$-form on the
moduli space, one can evaluate it on a basis of  tangent vectors using the
expression $dC^{\dagger}(\mu_1,
\mu_2)=\mu_1(C^{\dagger}(\mu_2))-\mu_2(C^{\dagger}(\mu_1))-C^{\dagger}([\mu_1,
\mu_2])$, and take into account that, as $\mu_1$, $\mu_2$ are constant vector
fields, their Lie bracket is $0$.  Finally, we must compute the section term
in \mqaj. It takes the form,
$$
\eqalign{ |s(A,M)|^{2}&={1 \over 2}\int
_{X}e (F^{+\alpha \beta} +{i\over 2}{\overline M}^{(\alpha}  M^{\beta )})
(F^{+}_{\alpha \beta} +{i\over 2}{\overline M}_{(\alpha}  M_{\beta )}) +
\int _{X} e {D}_{\alpha\dot\alpha} {\overline M}^{\alpha} D_{\beta}{}^{\dot
\alpha }M^{\beta} \cr & =\int _{X}e\big[ g^{ij}D_i\overline M^\alpha D_j
M_\alpha + {1\over 4} R \overline M^\alpha M_\alpha +{1\over 2}
F^{+\alpha\beta} F_{\alpha\beta}^+ -{1\over 8} \overline M^{(\alpha}
M^{\beta)}
      \overline M_{(\alpha} M_{\beta)} \big], \cr}
\eqn\melon
$$
where $R$ is the scalar curvature (not to be confused with the operator $R$
in \lima) on $X$. To obtain the second expression in this equation one can
either write explicitly the form of the Dirac operator, or, alternatively,
one can integrate by parts, use the relation $D_{\alpha\dot\alpha}
D_\beta{}^{\dot\alpha} M^\beta= (g^{ij}D_i D_j-{1\over 4} R) M_\alpha + i
F^+_{\alpha\beta} M^\beta$,  and then integrate back by parts. Notice that
if one denotes the components of $M_\alpha$ by $M_\alpha=(a,b)$, the last
factor in \melon\ is in fact ${1\over 2}(|a|^2 +|b|^2)^2$, and therefore it
is positive definite. The factor $i\overline M^\alpha F^+_{\alpha\beta}
M^\beta$ has cancelled in the sum, and then each term in the second
expression for $|s(A,M)|^2$ in \melon\ is positive definite except the one
involving the scalar curvature. This was the reason of choosing  the factor
${1/ {\sqrt 2}}$ in \rosa. The advantage of this form of the bosonic sector
in the action is that one can apply vanishing theorems which improve  the
analysis of the space of solutions of the monopole equations [\mfm, \wv].

The action resulting after adding all the terms in \fresa\ to \melon\ is
manifestly topological, for it is the field theoretical representation of
the Thom class of the bundle ${\cal E}$. Let us show how it can be obtained
in a more standard way from a BRST symmetry  ({\it i.e.} a nilpotent $Q$
operator up to gauge transformations) and an appropriate choice of gauge
fermion $\Psi$ such that the action resulting from \fresa\ and \melon\ is
$-\{Q, \Psi \}$ after introducing  auxiliary fields.  This approach to
topological quantum field theories  can be regarded from the traditional
BRST point of view initiated in \REF\labper{J.M.F. Labastida and M.
Pernici\journal\pl&B212(88)56} [\labper] and reviewed in \REF\blth{D.
Birmingham, M. Blau, M. Rakowski and G. Thompson\journal\pre&209(91)129}
[\blth], or from a modern perspective as described in [\cmrlect]. We will
follow in this paper the latter. In a topological field theory with gauge
symmetry there exists a localization gauge fermion which comes directly from
the Cartan model representative of the Thom class \mqaj\ with additional
auxiliary fields $(H,h)$,
$$
\Psi _{\rm loc}= -i\langle (\chi,
v),s(A,M) \rangle-{1 \over 4}\langle  (\chi, v),(H,h) \rangle,
\eqn\tigre
$$
and a projection gauge fermion which implements the horizontal
projection,
$$
\Psi _{\rm proj}=i{\langle \lambda , C^{\dagger} (\psi ,
\mu ) \rangle}_{g}.
\eqn\oso
$$

Using the $Q$-transformations \pera\ we obtain the  localization  and the
projection lagrangian, respectively:
$$
\eqalign{ \{Q,\Psi _{\rm loc}\} =
&\{ Q, \int _{X} e \big[-\chi^{\alpha \beta}\big({i \over {\sqrt 2}}
(F^{+}_{\alpha \beta} +{i \over 2}{\overline M}_{(\alpha}M_{\beta)} ) +{1
\over 4} H_{\alpha \beta} \big)\cr
&\,\,\,\,\,\,\,\,\,\,\,\,\,\,\,\,\,\,\,\, -{i\over 2} ({\bar v}^{\dot
\alpha}D_{\alpha \dot \alpha} M^{\alpha}+{\overline M}^{\alpha}D_{\alpha \dot
\alpha}v^{\dot \alpha})-{1 \over 8} ({\bar v}^{\dot \alpha} h_{\dot
\alpha}-{\bar h}_{\dot \alpha}v^{\dot \alpha} ) \big] \}\cr =  & \int _{X}e
\big[-{i \over {\sqrt 2}}
 H^{\alpha \beta} (F^{+}_{\alpha \beta} + {i\over 2}{\overline M}_{(\alpha}
M_{\beta )}) +{i \over {\sqrt 2}}\chi ^{\alpha
\beta}\big((p^+(d\psi))_{\alpha
\beta}+{i\over 2}({\bar \mu}_{(\alpha}M _{\beta)}+ {\overline
M}_{(\alpha}\mu _{\beta)}\big)\cr  &\,\,\,\,\,\,\,\,\,\, -{1 \over 4}
H^{\alpha \beta} H_{\alpha \beta} - {i\over 2} ({\bar h}^{\dot
\alpha}D_{\alpha \dot \alpha} M^{\alpha}+{\overline M}^{\alpha}D_{\alpha
\dot \alpha}h^{\dot \alpha})+{i \over 2}({\bar v}^{\dot \alpha}D_{\alpha
\dot \alpha} \mu^{\alpha}-{\bar  \mu}^{\alpha}D_{\alpha \dot \alpha}v^{\dot
\alpha})\cr  &\,\,\,\,\,\,\,\,\,\, +{1 \over 2} ({\overline M}^{\alpha}
\psi_{\alpha \dot \alpha}v^{\dot \alpha}- {\bar v}^{\dot \alpha}
\psi_{\alpha \dot \alpha}M^{\alpha})-{1 \over 4} ({\bar h}^{\dot\alpha}
h_{\dot\alpha}+i\phi {\bar v}^{\dot\alpha} v_{\dot\alpha}) \big],\cr}
\eqn\manza
$$
$$
\eqalign{ \{Q,\Psi _{\rm proj}\}= &\{ Q, -\int _{X} \big[
i \lambda \wedge *d^{*} \psi +{1 \over 2} e \lambda ({\bar \mu}^{\alpha}
M_{\alpha}- {\overline M}^{\alpha} \mu_{\alpha}) \big] \}\cr  =&-\int _{X}
\big[i\big( \eta \wedge *d^{*} \psi + \lambda \wedge *d^{*}d \phi \big) + {1
\over 2} e \eta ({\bar \mu}^{\alpha} M_{\alpha} - {\overline M }^{\alpha}
\mu_{\alpha})\cr  & \,\,\,\,\,\,\, - e \lambda ({\bar
\mu}^{\alpha}\mu_{\alpha} -i\phi {\overline M}^{\alpha}M_{\alpha} \big)\big].
\cr}
\eqn\limon
$$

The sum of \manza\ and \limon\ is just the same as the
sum of the terms in \fresa\ plus $-|s(A,M)|^2$ as given in \melon\
once the auxiliary fields
$H_{\alpha\beta}$ and $h_{\dot\alpha}$ have been integrated out.
This is indeed the exponent appearing in the Thom class \mqaj\ which must
be identified as minus the action, $-S$, of the topological quantum field
theory. This action turns out to be:
$$
\eqalign{
S&=\int _{X}e\big[ g^{ij}D_i\overline M^\alpha D_j M_\alpha + {1\over 4} R
\overline M^\alpha M_\alpha +{1\over 2} F^{+\alpha\beta} F_{\alpha\beta}^+
-{1\over 8} \overline M^{(\alpha} M^{\beta)} \overline M_{(\alpha}
M_{\beta)} \big] \cr
&+i\int_{X} \big( \lambda  \wedge *d^{*}d \phi - {1
\over {\sqrt2}}\chi \wedge * p^{+}d \psi +\eta\wedge *d \psi \big) \cr
&+\int_{X}e\Big(i\phi \lambda{\overline M}^{\alpha}M_{\alpha} +{1 \over
2{\sqrt2}} {\chi}^{\alpha \beta}({\overline M}_{(\alpha} \mu _{\beta )}+{\bar
\mu}_{(\alpha} M_{\beta )}) -{i\over 2} (v^{\dot \alpha} D_{\alpha
{\dot\alpha}} {\mu}^{\alpha}-{\bar \mu}^{\alpha} D_{\alpha
{\dot\alpha}}v^{\dot \alpha}) \cr &\,\,\,\,\,\,\,\,\,\,\,\,\,\, -{1 \over
2}({\overline M}^{\alpha}{\psi}_{\alpha {\dot \alpha}}v^{\dot \alpha}-{\bar
v}^{\dot \alpha}{\psi}_{\alpha {\dot \alpha}}M^{\alpha}) +{1\over2}  \eta
({\bar \mu} ^{\alpha} M_{\alpha}-{\overline M}^{\alpha} \mu_{\alpha}) +{i
\over 4}\phi {\bar v}^{\dot\alpha} v_{\dot\alpha} -\lambda {\bar
\mu}^{\alpha} \mu _{\alpha}\Big).\cr }
\eqn\action
$$
This action is
invariant under the modified BRST transformations which are obtained from
\pera\ after integrating out the auxiliary fields. It contains the standard
gauge fields of a twisted $N=2$ vector multiplet, or Donaldson-Witten
fields, coupled to the ``matter" fields of the twisted $N=2$ hypermultiplet.

The observables of the theory are built out of products of BRST invariant
operators which are cohomologically non-trivial. These observables are based
on forms which can be grouped into families labeled by a positive integer
$n$. These forms can be obtained solving the standard descent equations
[\tqft] or using the $G_i$ operators in [\alabas]. One can also use the
method explained in
\REF\bs{L. Baulieu and I.M. Singer, {\it Nucl. Phys. Proc. Suppl.} {\bf 5B}
(1988) 12}[\bs].
They turn out to be:
$$
\eqalign{ \Theta_0^n &= {n\choose 0}\phi^n,  \,\,\,\,\,\,\,\,\,\,\,\,\,\,
\Theta_1^n = {n\choose 1}\phi^{n-1}\psi, \cr \Theta_2^n &= {n\choose
2}\phi^{n-2}\psi\wedge\psi+{n\choose 1}\phi^{n-1}F, \cr \Theta_3^n &=
{n\choose 3}\phi^{n-3}\psi\wedge\psi\wedge\psi+
            2{n\choose 2}\phi^{n-2}\psi\wedge  F, \cr \Theta_4^n &=
{n\choose 4}\phi^{n-4}\psi\wedge\psi\wedge\psi\wedge\psi+
            3{n\choose 3}\phi^{n-3}\psi\wedge\psi \wedge F+
             {n\choose 2}\phi^{n-2}F\wedge F. \cr}
\eqn\forms
$$
The ghost number of the $i$-form $\Theta_i^n$  is $2n-i$ for $i=0,1,2,3,4$.
We have  not found non-trivial observables involving the ``matter" fields.
The forms  \forms\ verify the descent equations,
$$
[Q,\Theta_i^n\} =
d\Theta_{i-1}^n,
\eqn\descent
$$
and therefore one can assocaite a
$Q$-invariant operator to each of them in the following way. Let $x$ denote
a point in $X$, and $\gamma_j$ a $j$-cycle for $j=1,2,3$. The $Q$-invariant
operators have the form:
$$
\eqalign{ {\cal O}_0(n,x) &= \Theta_0^n(x),\cr
&\cr {\cal O}_1(n,\gamma_1) &= \int_{\gamma_1}\Theta_1^n,
\,\,\,\,\,\,\,\,\,\,\,\,\,\,\,\,\,\,\,\, {\cal O}_3(n,\gamma_3) =
\int_{\gamma_3}\Theta_3^n,\cr {\cal O}_2(n,\gamma_2) &=
\int_{\gamma_2}\Theta_2^n, \,\,\,\,\,\,\,\,\,\,\,\,\,\,\,\,\,\,\,\, {\cal
O}_4(n) = \int_X\Theta_4^n.\cr}
\eqn\operators
$$
Observables are built out
of products of these operators. In order to have non-trivial contributions
the ghost number of these products  must match the ghost-number anomaly in
the theory, which coincides with the index calculated in \leon. This is a
necessary condition to get a non-trivial vacuum expectation value but
certainly is not sufficient.

In [\mfm] Witten showed  that in certain situations in which the
ghost-number anomaly vanishes the sum of the partition function of the
theory over classes of $U(1)$ bundles such that the index in \leon\ vanishes
is related to Donaldson invariants. The construction of the topological
quantum field theory provides a rich set of operators whose vacuum
expectation values might lead to interesting topological invariants in more
general situations. Certainly, it opens the possibility of discovering new
topological invariants unrelated to Donaldson invariants.

The theory constructed in this work can be generalized in serveral
directions. One would correspond to the abelian $U(1)^n$ case with ``matter"
fields carrying different charges. This generalization is rather
straightforward after our construction for the simple $U(1)$ case.  More
interesting but certainly not so simple  is its non-abelian counterpart. The
construction of  the non-abelian generalization  can be carried out also in
the framework of the Mathai-Quillen formalism  using  technics which are
similar to the ones used in this paper.  Work in this direction will be
reported elsewhere.

\vskip1cm

\ack We would like to thank A. V. Ramallo for very helpful discussions.
M.M. would like to
thank G. Moore and S. Rangoolam for many conversations about the
Mathai-Quillen formalism.
This work was supported in part by DGICYT under grant PB93-0344 and
by CICYT under grant
AEN94-0928.

\vskip1cm

\refout \end